\documentclass{aa}  

\usepackage{xcolor}
\usepackage{soul}
\usepackage{url}
\usepackage{graphicx}
\usepackage{txfonts}
\usepackage{natbib}
\bibpunct{(}{)}{;}{a}{}{,}

\begin{document}

\title{The mean solar butterfly diagram\\ and poloidal field generation rate at the surface of the Sun}

\titlerunning{The mean solar butterfly diagram}
\authorrunning{S. Cloutier et al.}

\author{S. Cloutier$^1$,
          R. H. Cameron$^1$,
          \and
          L. Gizon$^{1,2}$
          }

\institute{Max-Planck-Institut f{\"u}r Sonnensystemforschung, Justus-von-Liebig-Weg 3, D-37077 G{\"o}ttingen, Germany\\
              \email{cloutier@mps.mpg.de} 
              \and
              Institut f\"ur Astrophysik und Geophysik, Georg-August-Universit\"at G\"ottingen, D-37077 G{\"o}ttingen, Germany}

\abstract
{The difference between individual solar cycles in the magnetic butterfly diagram can mostly be ascribed to the stochasticity of the emergence process.}
{We aim to obtain the expectation value of the butterfly diagram from observations of four cycles. This allows us to further determine the generation rate of the surface radial magnetic field.}
{We used data from Wilcox Solar Observatory to generate time-latitude diagrams of the surface radial and toroidal magnetic fields spanning cycles 21 to 24. We symmetrized them across the equator and cycle-averaged them. From the mean butterfly diagram and surface toroidal field, we then inferred the mean poloidal field generation rate at the surface of the Sun.}
{The averaging procedure removes realization noise from individual cycles. The amount of emerging flux required to account for the evolution of the surface radial field is found to match that provided by the observed surface toroidal field and Joy's law.} 
{Cycle-averaging butterfly diagrams removes realization noise and artefacts due to imperfect scale separation and corresponds to an ensemble average that can be interpreted in the mean-field framework. The result can then be directly compared to $\alpha\Omega$-type dynamo models. The Babcock-Leighton $\alpha$-effect is consistent with observations, a result that can be appreciated only if the observational data are averaged in some way.} 

               
\maketitle
%
%-------------------------------------------------------------------

\section{Introduction} \label{sect:intro}
The 11-year solar cycle is believed to be driven by a hydromagnetic dynamo seated somewhere in the convection zone of the Sun \citep{Charbonneau2020}. The dynamo loop consists of two parts: one where toroidal field is generated from the poloidal field, and the other where poloidal field of opposite sign is generated from the new toroidal field. It takes another solar cycle, or half of the magnetic cycle, to revert to the original polarity. The first part is relatively well understood and takes place via the $\Omega$-effect, whereby the poloidal field is wound up in the azimuthal direction by differential rotation, which is observationally well constrained by helioseismology \citep[e.g.][]{Schou1998}. The \mbox{second} part, on the other hand, involves the unobservable subsurface toroidal field and two possible mechanisms: the $\alpha$-effect \citep{Parker1955,Steenbeck1966} and the Babcock-Leighton (BL) mechanism \citep{Babcock1961,Leighton1964,Leighton1969}. This dynamo loop is often studied using mean-field models.

The mean field of mean-field electrodynamics is defined in terms of an average that can be spatial, temporal, or ensemble. In the case of the Sun, the azimuthal average is a natural choice. In fact, the butterfly diagram is constructed by longitudinally averaging line-of-sight synoptic magnetograms. The symmetric component (with respect to the solar meridian) is then extracted, from which the radial field is obtained by assuming it is the only component of the poloidal field at the surface. Finally, the butterfly diagram is obtained by stacking the result in time. In principle, butterfly diagrams are thus of a mean-field nature and can be directly compared to models. 

A caveat, however, is that scale separation is likely to be poor in the Sun as supergranulation reaches down to about 5\% of the solar radius, which is non-negligible with respect to the depth of the convection zone and the scales of differential rotation and the meridional flow. The picture is much worse if we consider active regions (which, by being largely non-axisymmetric, are supposed to be small-scale in nature), with sizes of the order of 100~Mm, or about half the depth of the convection zone. Consequently, the observed butterfly diagram cannot be directly compared to mean-field models. A way to circumvent this problem is by considering different solar cycles as different realizations of the underlying dynamo and ensemble-averaging them; we note that this is not a true ensemble average in the mean-field sense (see \citealt{Hoyng2003}) but is aimed at reproducing the azimuthal \mbox{average} used in models with reduced realization noise. In other words, averaging different cycles together to remove realization noise (fluctuating dynamo parameters) and the effects of imperfect scale separation.

\begin{figure}
\resizebox{\hsize}{!}{\includegraphics{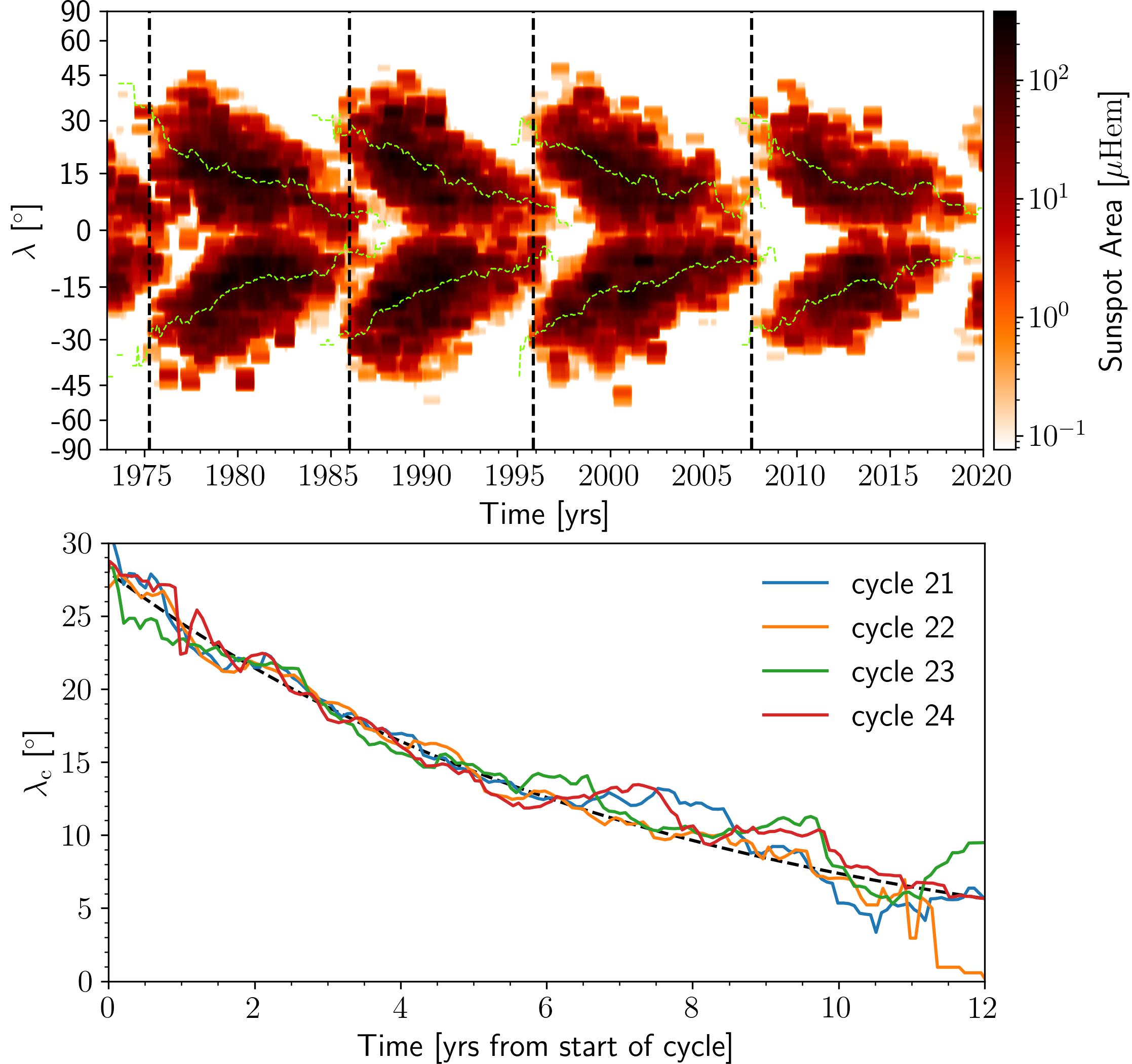}}
\caption{12-month average of the sunspot area (top) and centroid latitudes as a function of time starting from $t_0$ (bottom). The dashed green lines represent the centroid latitude of the sunspot zone (Eq. (\ref{eq:centroid})) and the vertical dashed lines the reference start times of the cycles (see text).}
 \label{fig:area}
\end{figure}

The way cycles are to be averaged together requires careful consideration. It has been shown that defining cycles in terms of activity minimum or maximum is not ideal \citep{Waldmeier1935, Hathaway2011}. But when cycles are instead defined according to their start time, they all present the same equatorward \mbox{pattern} of sunspot migration, regardless of strength; the location of the sunspot area centroid migrates towards the equator according to the same standard law \citep{Waldmeier1939,Waldmeier1955,Hathaway2011,CS2023}. The individual cycles can then be averaged together in phase. To carry out this cycle-averaging, we used butterfly diagrams produced from synoptic magnetograms obtained at Wilcox Solar Observatory (WSO). We further symmetrized the data across the equator so that we have, in effect, an average consisting of eight cycles.

In addition to the surface radial field, another important quantity we can infer from longitudinally averaged line-of-sight magnetograms is the toroidal field emerging at the surface, the antisymmetric component (with respect to the solar meridian), from which we can construct a time-latitude diagram \citep[from WSO data;][]{Cameron2018}. Having mean time-latitude \mbox{diagrams} of both the surface radial and toroidal fields at our disposal, we can test the validity of the simple $\alpha\Omega$ dynamo models, as well as possibly constrain the functional form of the source term. To do this, we used both quantities to independently calculate an intermediate one: the surface radial field generation rate (i.e. the radial source term). We inferred this quantity from the butterfly diagram using the surface flux transport (SFT) model \citep{Leighton1964,DeVore1984,Sheeley1985}, and from the time-latitude diagram of the surface toroidal field using Joy's law. We then compared the two diagrams to determine if they are consistent with each other. 

\begin{figure}
\resizebox{\hsize}{!}{\includegraphics{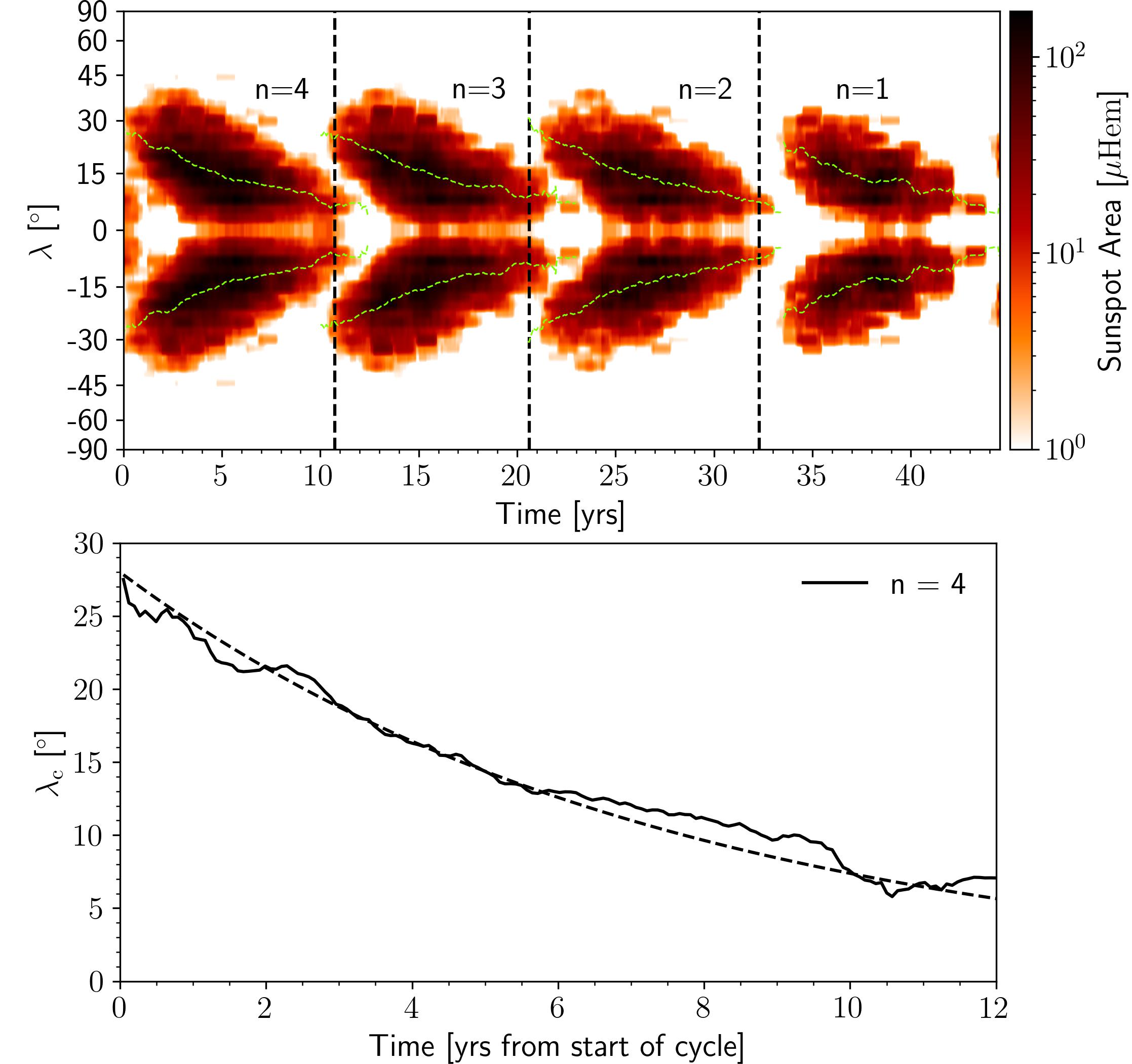}}
\caption{Cycle average of the sunspot area (top) and centroid latitudes as a function of time starting from $t_0$ (bottom). The dashed green lines represent the centroid latitude of the sunspot zone (Eq. (\ref{eq:centroid})) and the vertical dashed lines the reference start times of the cycles (see the main text).}
 \label{fig:area_avg}
\end{figure}

\section{The equatorward migration of the sunspot zones} \label{sect:law}

The original work of \citet{Waldmeier1939,Waldmeier1955} concerning the equatorward migration of the sunspot zones was extended by \citet{Hathaway2011}, who used the then entire daily sunspot area record (cycles 12 to 22) as provided by the Royal Observatory, Greenwich (May 1874 to December 1976) and the United States Air Force and National Oceanic and Atmospheric Administration (from January 1977). The average daily sunspot area for each Carrington rotation, $A(\lambda_i)$, was then calculated, as was the latitude of the centroid of the sunspot area for each hemisphere and Carrington rotation, separately for each cycle, according to a sunspot area weighted average,
\begin{equation}
    \lambda_\mathrm{c}=\frac{\sum_i A(\lambda_i)\lambda_i}{\sum_i A(\lambda_i)}, \label{eq:centroid}
\end{equation}
the sums being carried out over the 25 latitude bins of each hemisphere. The start time of each cycle, $t_0$, was then determined by fitting the monthly sunspot number record of each cycle to a skewed Gaussian devised by \citet{Hathaway1994}. This procedure allowed \citet{Hathaway2011} to confirm the results of \citet{Waldmeier1939,Waldmeier1955}, that the latitudinal drift of the sunspot zones follows the same universal path, regardless of \mbox{cycle} strength or phase. This standard law was found to closely approximate the following decaying exponential:
\begin{equation}
    \lambda_\mathrm{c}(t)=28^\circ\exp{\left(-\frac{t-t_0}{90~\textrm{mths}}\right)}. \label{eq:law}
\end{equation}

Our cycle-averaging procedure requires the determination of $t_0$ for each cycle. To do so, we first took the 12-month average of the sunspot area, only considering the four cycles for which we have magnetograms (Fig. \ref{fig:area}).\footnote{Downloaded from \url{http://solarcyclescience.com/activeregions.html}.} We then determined the centroid latitudes of the sunspot belts by applying Eq. (\ref{eq:centroid}), averaged both hemispheres together, and lastly fitted the first six years of these curves with Eq. (\ref{eq:law}). The result is shown in Fig. \ref{fig:area}. 

We considered the average of the quantity $f$ defined as 
\begin{equation}
    \overline{f}(\theta,t-t_{01})=\sum_{i=1}^n\frac{f(\theta,t-t_{0i})}{n_k(t)},\label{eq:mean}
\end{equation}
$n$ being the number of cycles that are averaged and $n_k(t)=n+i-k$ for $t_{0k}\leq t <t_{0k+1}$. The cycle-averaging was performed as \mbox{follows}: four copies of the data were made; then for each copy, $i$, the whole dataset was shifted back so that the cycle start times $t_{0i}$ and $t_{01}$ coincide, whereafter the shifted copies were averaged together. We further symmetrized the data across the equator. Our procedure explicitly minimizes any north-south \mbox{asymmetry}, which will also contain information about the dynamo. The result for the 12-month average of the sunspot area is shown in Fig. \ref{fig:area_avg}. It consists of four averages of the solar cycles obtained by including a decreasing amount of cycles in the average.

\section{The observed mean butterfly diagram}

\begin{figure}
\resizebox{\hsize}{!}{\includegraphics{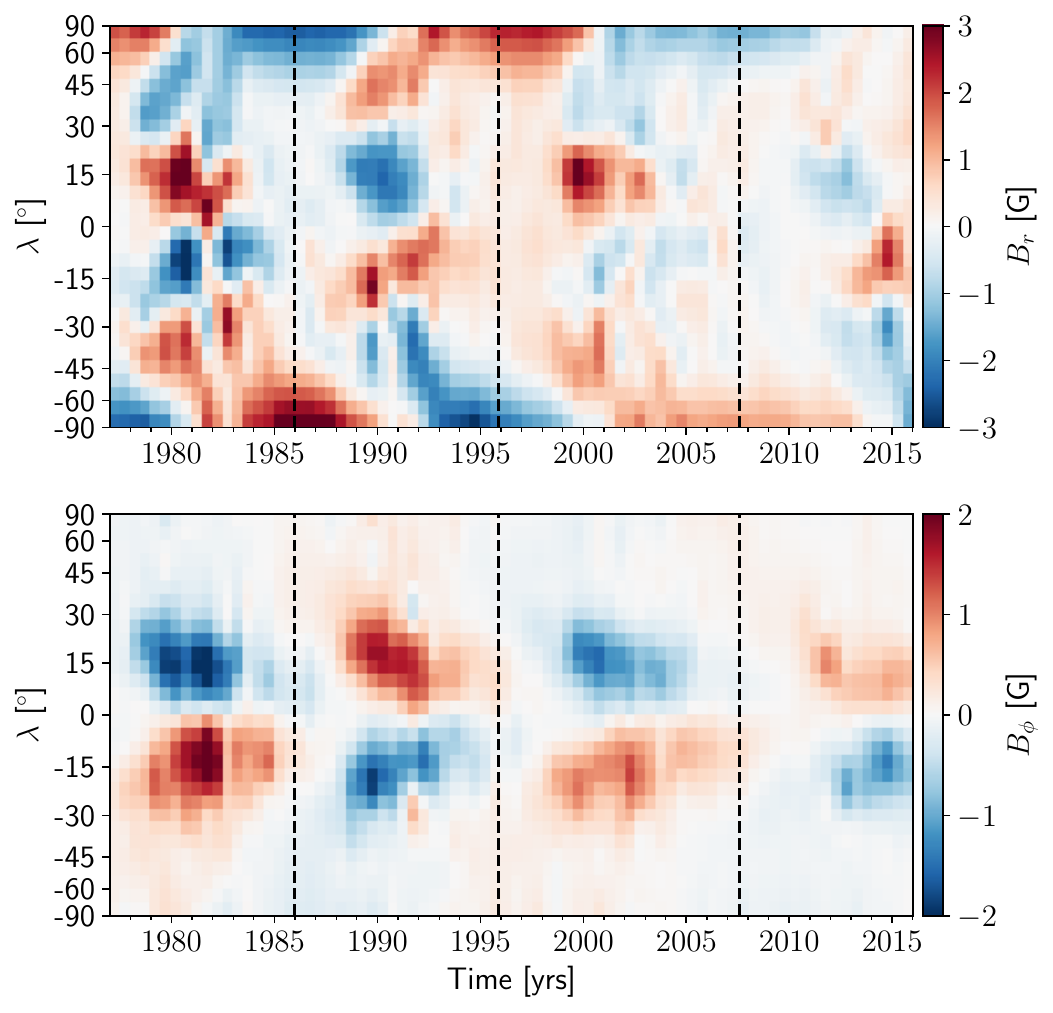}}
\caption{Time-latitude diagrams of the surface radial field, $B_r$ (top), and surface toroidal field, $B_\phi$ (bottom). The vertical dotted lines indicate the cycle start times as determined by the procedure outlined in Sect. \ref{sect:law}.}
 \label{fig:butt_obs}
\end{figure}

\begin{figure}
\resizebox{\hsize}{!}{\includegraphics{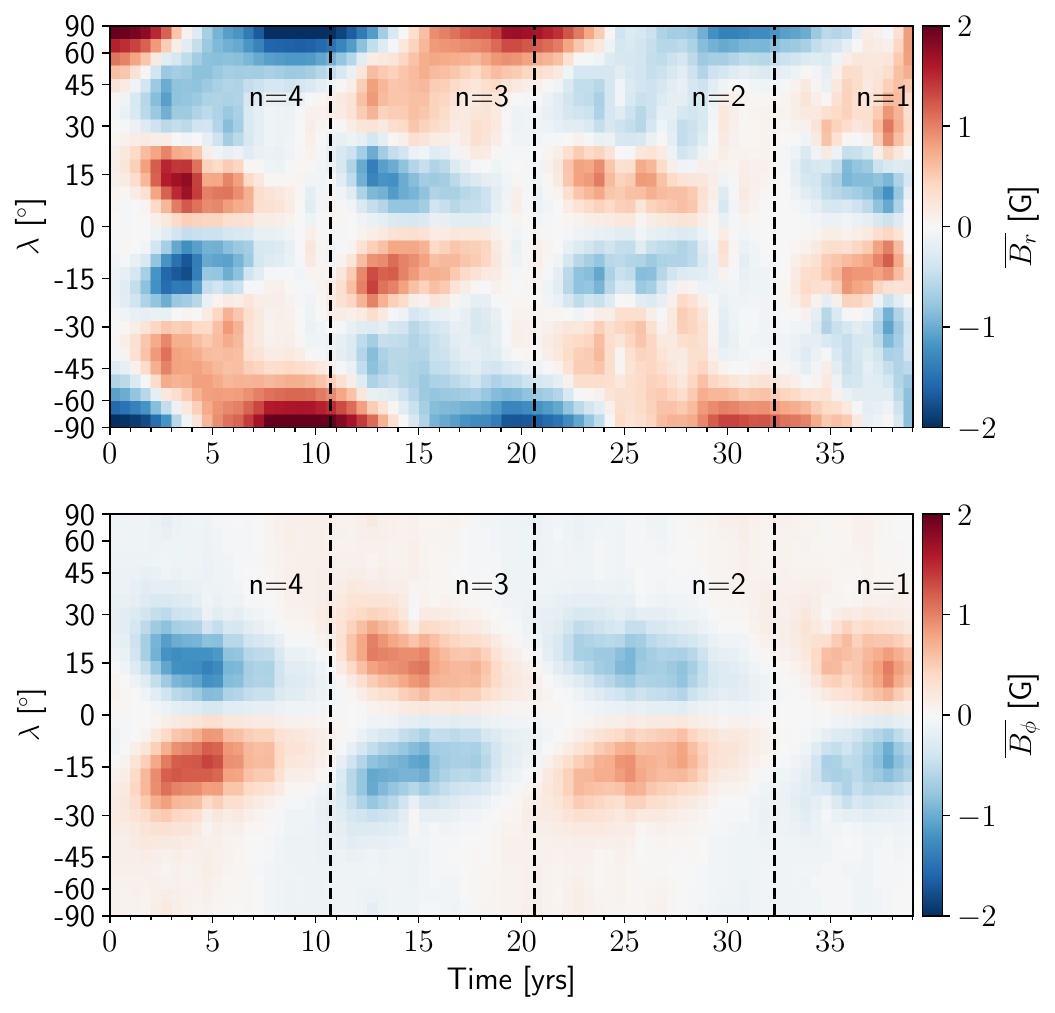}}
\caption{Time-latitude diagrams of the mean surface radial field, $\overline{B_r}$ (top), and surface toroidal field, $\overline{B_\phi}$ (bottom). The vertical dotted lines indicate the cycle start times as determined by the procedure outlined in Sect. \ref{sect:law}.}
 \label{fig:butt_obs_avg}
\end{figure}

The WSO data we used are presented in Fig. \ref{fig:butt_obs} along with the reference start times of the individual cycles (see Sect. \ref{sect:law}). By carrying out our cycle-averaging procedure, we obtained the time-latitude diagrams presented in Fig. \ref{fig:butt_obs_avg}. One can immediately see that the cycle-averaging produces butterfly wings that are much better defined. We have clear butterfly wings of one polarity, that of the leading sunspots, and poleward \mbox{migrating} fields of the opposite polarity, that of the trailing sunspots. The latter feature is also known as the `rush to the poles' \mbox{\citep{Ananthakrishnan1954,Altrock1997}}. Even with only eight \mbox{realizations}, the trailing polarity field is essentially removed from the butterfly wings. The rush is also increasingly filled with fields of trailing polarity and is devoid of fields of leading polarity \citep[see Fig. 3 of][for a more detailed look at the small-scale surges and features]{Mordvinov2022}. We thus see a `\mbox{separation} of \mbox{polarities}' happening. The butterfly wings are much more clearly delineated and seem narrower, with a width slightly under $30^\circ$ rather than $\simeq 35^\circ$. Lastly, the rapid increase in the sunspot number during the rising phase, and its slow decrease during the declining phase, is much more apparent.

It is clear, however, that four solar cycles is not nearly enough to carry out a proper ensemble average. It is nonetheless reasonable to assume that the addition of more cycles would only further the observed trend. Most of the small-scale features will be washed out and the mean observed butterfly diagram will look very similar to those calculated from mean-field dynamo models: smooth butterfly wings of the leading spot polarity, and a rush to the poles and polar fields of the trailing spot polarity.

We note the change of the scaling from Fig. \ref{fig:butt_obs} to \ref{fig:butt_obs_avg}: the largest fields inside the butterfly wings are reduced from 3 G to 2 G. Higher-resolution butterfly diagrams \citep[e.g.][]{Norton2023} have fields of the order of 10 G; in WSO data, part of the averaging has already been done by the instrument's limited resolution. This is more evidence that one needs to be careful when comparing simulation results from mean-field models to observations.

\section{Inferring the radial field generation rate}
The radial field generation rate, or the radial source term, can be inferred both through the butterfly diagram and the time-latitude diagram of the surface toroidal field by using SFT and emergence (BL mechanism) models.

\subsection{Surface flux transport model} \label{sect:sft}
The SFT model describes the passive advective and diffusive transport of the surface radial field via the following equation \citep{Yeates2023}:
\begin{equation}
\begin{aligned}
    \frac{\partial B_r}{\partial t}=&-\omega(\lambda)\frac{\partial B_r}{\partial\phi}-\frac{1}{R_\odot\cos\lambda}\frac{\partial}{\partial\lambda}\left(\cos\lambda v(\lambda)B_r\right)\\
    &+\frac{\eta}{R_\odot^2}\left[\frac{1}{\cos\lambda}\frac{\partial}{\partial\lambda}\left(\cos\lambda\frac{\partial B_r}{\partial\lambda}\right)+\frac{1}{\cos^2\lambda}\frac{\partial^2B_r}{\partial\phi^2}\right]+S_r(\lambda,\phi,t),
\end{aligned}
\end{equation}
where $B_r$, $\omega(\lambda)$, $v(\lambda)$, $\eta$, and $S_r(\lambda,\phi,t)$ are respectively the surface radial field, angular velocity, meridional flow, diffusivity, and radial source term; $\lambda$ is the latitude and $\phi$ the longitude. Since the magnetic butterfly diagram provides us with the longitudinally averaged surface radial field, $B_r(\lambda,t)$, the only unknown is the surface radial source term,
\begin{equation}
\begin{aligned}
    S_r(\lambda,t)=&~\frac{\partial B_r}{\partial t}+\frac{1}{R_\odot}\frac{\partial}{\partial\sin\lambda}\left(\cos\lambda v(\lambda)B_r\right)\\
    &-\frac{\eta}{R_\odot^2}\frac{\partial}{\partial\sin\lambda}\left(\cos^2\lambda\frac{\partial B_r}{\partial\sin\lambda}\right).
\end{aligned}
\end{equation}
We used a value of turbulent diffusivity of $\eta=350~$km$^2$/s, which is consistent with estimates from observations \citep[e.g.][]{Komm1995}, the SFT model from \citet{Lemerle2015}, and mixing length theory \citep[e.g.][]{MJ2011}. As for the surface meridional flow profile, we used the anti-symmetrized time-dependent profiles obtained from the helioseismic inversions of \citet{Gizon2020}.

\subsection{Emergence model} \label{sect:emer}
There is strong evidence \citep[see e.g.][]{DE2010,KO2011,CS2015} that the BL mechanism is the main poloidal field generation mechanism of the solar dynamo. The first part of this mechanism is the emergence through the photosphere of tilted toroidal flux tubes. The corresponding generation of radial field from the emerging toroidal field is derived as follows.

\begin{figure}
\resizebox{\hsize}{!}{\includegraphics{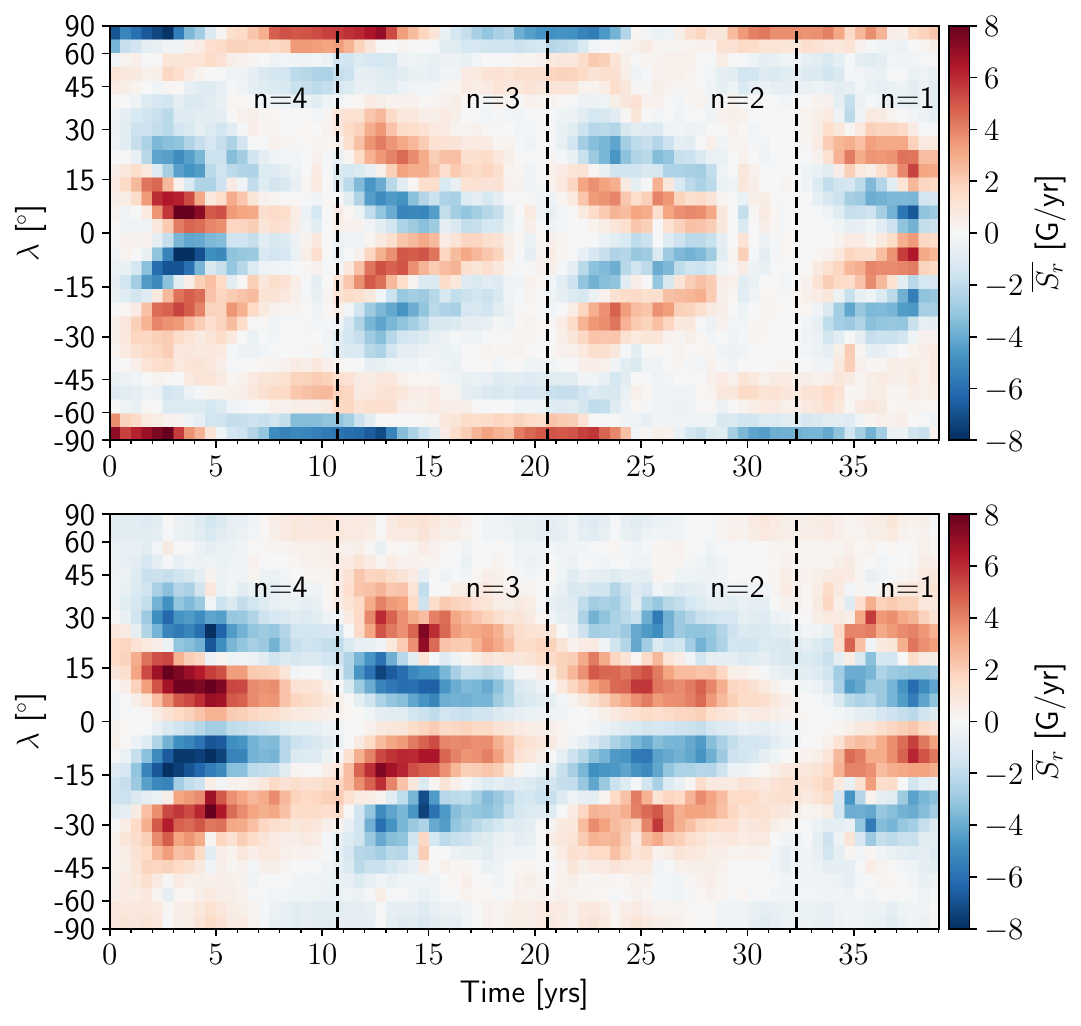}}
\caption{Time-latitude diagrams of the mean surface radial source term, $\overline{S_r}$, inferred from the surface radial field and SFT model (top) and from the surface toroidal field and emergence model for $m=0$ (bottom); see the main text. The vertical dotted lines indicate the cycle start times as determined by the procedure outlined in Sect. \ref{sect:law}.}
\label{fig:sr_obs_avg}
\end{figure}

\begin{figure}
\resizebox{\hsize}{!}{\includegraphics{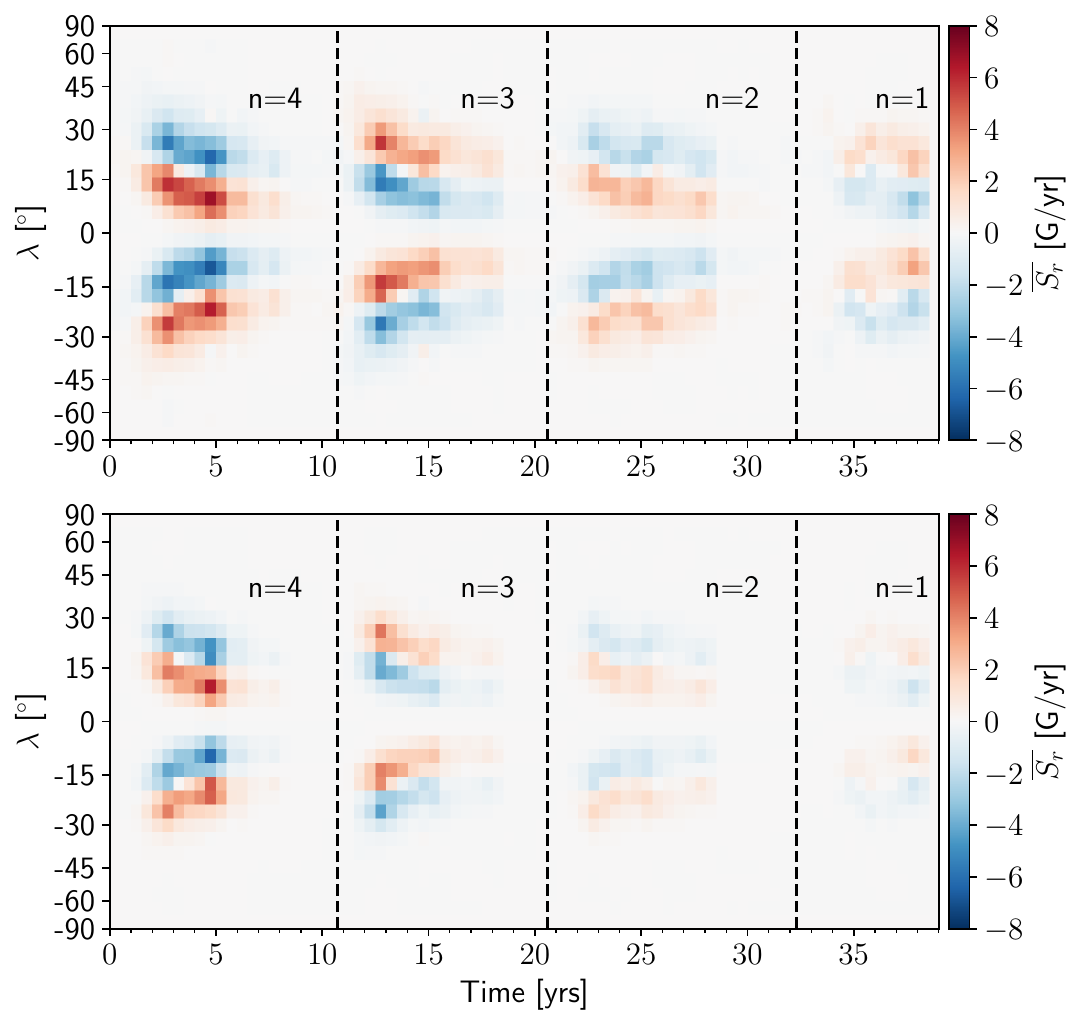}}
\caption{Time-latitude diagrams of the mean surface radial source term, $\overline{S_r}$, inferred from the surface toroidal field and emergence model for $m=1$ (top) and $m=2$ (bottom); see the main text. The vertical dotted lines indicate the cycle start times as determined by the procedure outlined in Sect. \ref{sect:law}.}
\label{fig:sr_obs_v}
\end{figure}

We begin with the solenoidal condition $\nabla \cdot {\boldsymbol B}=0$ for an axisymmetric field, ${\boldsymbol B}$,
\begin{eqnarray}
\frac{\partial(r^2 B_r)}{\partial r}
+\frac{r}{\sin \theta} \frac{\partial (B_\theta \sin \theta)}{\partial \theta}=0.
\end{eqnarray}
Next, we integrate in $r$ from the surface to infinity and take the time derivative to obtain 
\begin{eqnarray}
\frac{\partial B_r|_{R\odot}}{\partial t}
=-\frac{1}{R_\odot^2} \int_{R_\odot}^{\infty} \frac{r}{\sin \theta} \frac{\partial^2 (B_\theta \sin \theta) }{
\partial \theta \partial t} \mathrm{d} r.
\end{eqnarray}
Joy's law implies that as flux is emerging through the photosphere,
$B_\theta=B_\phi \tan \delta$, where $\delta$ is the tilt angle of the resulting bipolar magnetic region (BMR). The rate at which the flux of $B_\phi$ is carried through the surface per radian in $\theta$ is
\begin{equation}
\begin{aligned}
\frac{\partial}{\partial t}  \int_{R_\odot}^{\infty} B_\phi r \mathrm{d} r
=&- R_\odot v_e B_\phi|_{R_\odot},
\end{aligned}
\end{equation}
where $v_e$ is the radial velocity of the emerging field. Thus,
\begin{equation}
\begin{aligned}
\frac{\partial}{\partial t}  \int_{R_\odot}^{\infty} B_\theta r \mathrm{d} r=&~\tan\delta\frac{\partial}{\partial t}\int_{R_\odot}^{\infty}B_\phi r\mathrm{d}r,\\
=&-\tan\delta~ R_\odot v_e B_\phi.
\end{aligned}
\end{equation}
This yields, for the surface radial field generation rate,
\begin{equation}
\frac{\partial B_r|_{R\odot}}{\partial t}
= \frac{1}{R_\odot \sin \theta} \frac{\partial}{\partial\theta}\left(\sin\theta\tan\delta~ v_e B_\phi\right),
\end{equation}
or, with $\cos\delta\sim 1$,
\begin{equation}
    S_r(\lambda,t)=-\frac{1}{R_\odot}\frac{\partial}{\partial\sin\lambda}\left(\cos\lambda\sin\delta~ v_e B_\phi\right).
\end{equation}

We used the form of Joy's law as given by \citet{Leighton1969}, $\sin\delta=0.5\sin\lambda$. As for the emergence velocity, we considered three different prescriptions. The first is the constant velocity of $v_e\simeq200~$m/s inferred by \citet{Centeno2012}; we also considered the cases where the emergence velocity, $v_e$, is proportional to the toroidal field, $B_\phi$ \citep{Parker1975} or to its square, $B_\phi^2$ \citep{UR1976}, mimicking magnetic buoyancy \citep{Parker1955b,Jensen1955}. We thus write the emergence velocity as
\begin{equation}
    v_e=\left[\frac{|B_\phi|}{\max(B_\phi)}\right]^m v_0,
\end{equation}
where $v_0=200~$m/s and $m\in\{0,1,2\}$.

\section{The observed mean radial field generation rate}

The mean surface radial field generation rates, $\overline{S_r}$, as inferred from the surface radial (Sect. \ref{sect:sft}) and toroidal fields (Sect. \ref{sect:emer}), are shown in Fig. \ref{fig:sr_obs_avg}. We clearly see one `mean BMR' inside the activity belts. Remarkably, the two methods used to infer the radial field generation rate give results that are in good qualitative agreement, including for both the shape of the mean BMRs and their amplitudes. There are, however, three main noticeable discrepancies. Firstly, the SFT model gives us important radial field generation at the poles, of the opposite polarity to that of the polar fields. This could be explained by underestimated polar field strengths from the observations \citep{Linker2017}, possibly in combination with a weakened meridional flow amplitude at latitudes above $60^\circ$ \citep{Mahajan2021}. Secondly, there are sorts of weak-amplitude `bridges' at cycle minima between the poleward polarity of one cycle and the equatorward polarity of the next (showcasing the magnetic cycle and activity overlap between cycles). This overlap is only  visible in the $\overline{S_r}$ obtained through the surface toroidal field and Joy's law. Thirdly, the source terms derived from  the surface toroidal field and Joy's law are slightly wider and are shifted slightly polewards compared to those obtained from the surface radial field. 

Lastly, we assumed that the emergence velocities of the toroidal field depend linearly and quadratically on the toroidal field strength. The result is presented in Fig. \ref{fig:sr_obs_v}. We note that the cycle average was carried out after determining the surface radial source term. We see that increasing this dependence restricts the mean BMR to increasingly lower latitudes. Furthermore, in both the linear and quadratic cases, the aforementioned `bridges' between sunspot cycles seen in the constant velocity case disappear. Since the bridges are absent in the $\overline{S_r}$ time-latitude diagram necessary to explain the evolution of the surface radial field, this suggests that either Joy's law or the emergence velocity of low flux regions might differ from that of active regions.

\section{Conclusions}

In this paper we have shown that one must be careful when comparing butterfly diagrams produced by simple mean-field models with observed ones. Because of imperfect scale separation, the use of the azimuthal average does not result in an observed butterfly diagram that can be understood in the mean-field sense. The observed butterfly diagram must for this purpose be cycle-averaged, a procedure made possible by the universality of the equatorward sunspot belt migration. From this standard law we can determine a reference time when all cycles are in phase with each other and can thus be meaningfully averaged. To `increase' the small number of cycles for which magnetogram observations exist by a factor of two, we symmetrized both hemispheres together.

By making use of the average toroidal field time-latitude \mbox{diagram} and Joy's law, we found that the $\alpha$-effect-like BL source term is qualitatively consistent in shape and amplitude with the source term as inferred from the SFT model and the butterfly diagram. This consistency is further increased when we assume the toroidal field emergence velocity to be dependent on the toroidal field strength itself.

\begin{acknowledgements}
The authors wish to thank the anonymous referee for comments that helped improve the overall quality of this paper. This work was carried out when SC was a member of the International Max Planck Research School for Solar System Science at the University of Göttingen. The authors acknowledge partial support from ERC Synergy grant WHOLE SUN 810218. 
\end{acknowledgements}

\bibliographystyle{aa}
\bibliography{aa50739-24corr}

\end{document}